\documentclass[aps,prl,twocolumn,showpacs]{revtex4}
\usepackage{graphicx}
\usepackage{dcolumn}
\usepackage{bm}
\begin{document}
\preprint{cond-mat/0706.1268}
\title{Tunneling between Dilute GaAs Hole Layers}
\author{S. Misra}
\author{N. C. Bishop}
\author{E. Tutuc}
 \altaffiliation{present address: Microelectronics Research Center, University of Texas, Austin}
\author{M. Shayegan}
\affiliation{
Department of Electrical Engineering, Princeton University, Princeton, NJ 08544
}

\date{\today}

\begin{abstract}

We report interlayer tunneling measurements between very dilute
two-dimensional GaAs hole layers. Surprisingly, the
shape and temperature-dependence of the tunneling
spectrum can be explained with a Fermi liquid-based
tunneling model, but the peak amplitude is much larger
than expected from the available hole band parameters. Data as a function of
parallel magnetic field reveal additional anomalous features,
including a recurrence of a zero-bias tunneling peak at very large
fields. In a perpendicular magnetic field, we observe a robust
and narrow tunneling peak at total filling factor $\nu_T=1$, signaling
the formation of a bilayer quantum Hall ferromagnet.

\end{abstract}

\maketitle

Measurements of tunneling between a pair of low-disorder, GaAs
two-dimensional electron systems (2DESs) have revealed some of the
most basic \cite{EisenFS, EisenCoul, Katayama, Murphy, Ritchie, Spielman}, as well as
exotic \cite{Ian} properties of these systems. The
measurements have probed, e.g., the Fermi contour of the 2D
electrons \cite{EisenFS}, and have determined their quantum lifetime
\cite{Katayama, Murphy, Ritchie, Spielman, Macdonald1, Macdonald2, DS, ML, RW}. More
recently, tunneling experiments on closely spaced, interacting electron bilayers
at the total Landau level filling factor $\nu_T=1$ revealed an
unusual, excitonic quantum Hall ground state \cite{Ian}.

Here we report an experimental study of the differential
tunneling conductance between very dilute, closely spaced GaAs
2D {\it hole} systems (2DHS) as a function of interlayer bias
\cite{MM}. In each of the layers, the inter-hole
distance measured in units of the effective Bohr radius, $r_s=
{{m^*e^2}\over{4\pi\epsilon \epsilon_0 \hbar^2 \sqrt{\pi p}}}$, reaches up to $\simeq
13$, attesting to the extreme diluteness of the 2DHS (here, $p$ is
the 2D hole density in each layer, $m^* =0.2m_0$ is the
hole effective mass \cite{Mass}, and $\epsilon=12.4$ is the GaAs dielectric
constant). The physics of such a dilute system is expected to be
dominated by interaction effects. Indeed, in the presence of a
perpendicular magnetic field ($B_\perp$) and at total Landau level filling
factor $\nu_T=1$, we observe a strong zero-bias tunneling peak which
persists up to a temperature of $T\simeq 0.6$K, indicative of
significant interlayer {\it and} intralayer interaction in this
system \cite{Ian}. Despite the extreme diluteness of our bilayer 2DHS, the shape and
temperature-dependence of the tunneling peak in
{\it zero} applied magnetic field are consistent with an
existing model based on Fermi liquid theory \cite{Macdonald1, Macdonald2}.
The peak amplitude, however, is roughly 8 orders of magnitude larger than that calucated using
the conventional hole band parameters. Our measurements in samples with different barrier thicknesses
suggest that the hole mass relevant to tunneling through an AlAs barrier is much smaller than expected. Finally, tunneling measurements
performed with an applied parallel magnetic field $B_{||}$
reveal another puzzle: while the amplitude of the
zero-bias tunneling peak quickly decays with $B_{||}$ consistent
with the expected momentum conservation requirement, the peak reappears at very
large $B_{||}$.

We performed interlayer tunneling measurements on
double quantum well samples grown on GaAs
(311)A substrates. The structures consist of a
pair of 15 nm-wide GaAs quantum wells, seperated by 7.5, 8.0, or 11 nm AlAs
barriers, and flanked by Si-modulation-doped layers of Al$_{0.21}$Ga$_{0.79}$As.
While we will discuss the data from all three samples,
the data shown in this report are all from the sample with a 7.5 nm barrier.
The as-grown density for this sample is $\simeq 2\times10^{10}$
per layer, and the typical mobility at 0.3 K is  $\simeq 20$ m$^2/$Vs. A Y-shaped mesa,
illustrated schematically in the inset to Fig. 1(a), with an
active region of 100 $\mu$m $\times$ 700 $\mu$m, was etched into the
wafer. Alloyed InZn contacts at the ends of the seven arms allow for
making independent contacts to individual 2D hole layers using the
selective gate depletion scheme \cite{Eisencontacts}.
This Y-configuration enables us to measure
electrical transport in each of the two layers seperately, and to
balance their densities using front and back gates which cover the
active area, before proceeding with the
interlayer tunneling measurements. The latter were performed by
applying a low-frequency, $18 \mu$V peak-to-peak {\it ac} excitation on
top of an interlayer {\it dc} bias to the top layer, and measuring the
current in the back layer using standard lock-in techniques. The
lock-in phase was set by confirming that the measured differential
({\it ac}) conductance matched the derivative of the {\it dc} $I-V$ curve for
the tunneling measurement.

\begin{figure*}
\includegraphics{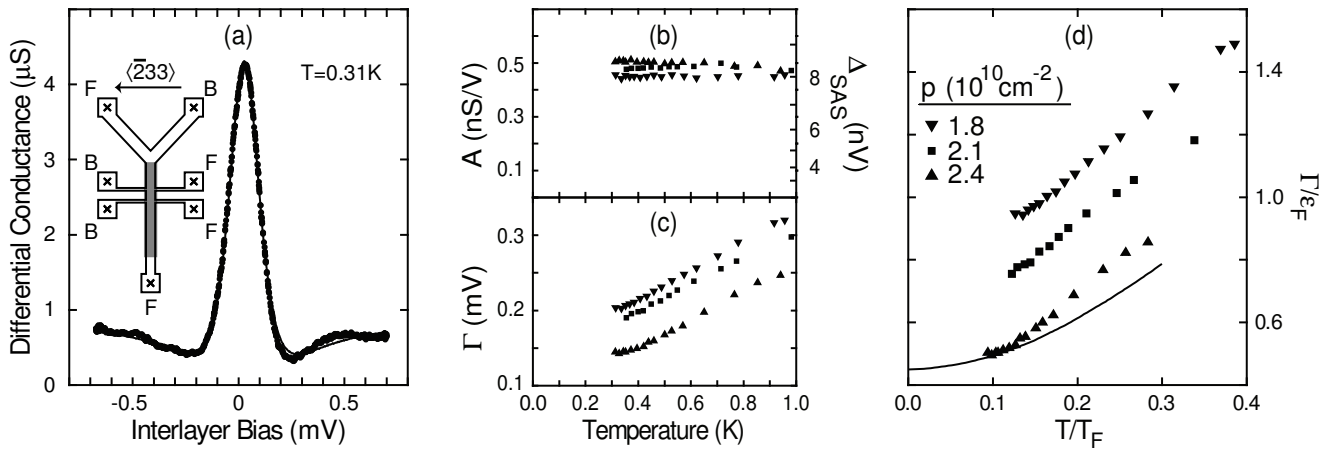}
\caption{\label{Fig1} (a) The differential conductance of the bilayer
hole system as a function of interlayer bias. The layers have equal
densities of $p=2.4\times10^{10}$cm$^{-2}$, corresponding to an
$r_s$ of 11, and are seperated by a 7.5 nm thick AlAs barrier.
The solid line represents a fit of Eq. (1)  to the data.
The inset shows a schematic diagram of our Hall bar. Front and back
gates cover the active (gray) area to allow for tuning the density
of each layer. Contacts (crosses) can be made to either
the front (F) or back (B) layer by energizing selective, depletion (contact)
gates (not shown) on the back or front of the sample, respectively.
Tunneling measurements were
performed by energizing all the contact gates, and using only the
four contacts in the middle of the Hall bar. (b \& c) Tunneling amplitude ($A$) and width ($\Gamma$),
derived from fitting the data to Eq. (1), are shown as a
function of temperature for three layer densities:
$p=1.8\times10^{10}$cm$^{-2}$ ($r_s=13$), $2.1\times10^{10}$cm$^{-2}$
($r_s=12$), and $2.4\times10^{10}$cm$^{-2}$ ($r_s=11$). Definition of symbols is provided in the inset to (d). (d) 
The width $\Gamma$, scaled to the Fermi energy ($\epsilon_F$),
is shown as a function of the temperature $T$, scaled to the Fermi
temperature ($T_F$). The solid line is the theoretical
prediction from Fig. 8 of Ref. \onlinecite{Macdonald2}, offset
vertically to account for the disorder scattering in our sample
at $p=2.4\times10^{10}$cm$^{-2}$.}
\end{figure*}


As illustrated in Fig. 1(a), the differential conductance of the tunnel
junction formed by the pair of 2D hole sheets with equal carrier
density shows a pronounced resonance near zero interlayer bias. This
zero-bias resonance is remarkably similar to that seen previously in
bilayer {\it electron} samples \cite{EisenCoul, Katayama, Murphy, Ritchie,Spielman},
despite here being taken in the extremely dilute limit ($r_s
\simeq 11-13$ for holes, $r_s\simeq 1-6$ for electrons). In the earlier works
\cite{Murphy, Ritchie, Spielman}, the $\it dc$ conductance, $I/V$, derived from the
data was found to closely match a theoretically derived \cite{Macdonald1, Macdonald2} expression:
${{I}\over{V}}(V)=St^2{{2e}\over{\hbar}}{{m^*}\over{\pi\hbar^2}}{{\Gamma}\over{(eV)^2+\Gamma^2}}$,
where $S$ is the area of the tunnel junction, $t$ is the tunneling
matrix element, $m^*$ is the density-of-states effective mass for the
2D carriers, $\Gamma$ is inversely proportional to the quasiparticle
scattering lifetime, and both $\Gamma$ and $|eV|$ are assumed to be
much smaller than the Fermi energy $E_F$. This zero-bias resonance
results from a close matching of the momenta and energies of the carriers
in the two wells. In Fig. 1(a), we
show a fit of our data to the differential conductance derived from
this expression \cite{Background2},
\begin{equation}
{{dI}\over{dV}}(V)=A\Gamma{{\Gamma^2-(eV)^2}\over{[\Gamma^2+(eV)^2]^2}}
\label{exp}
\end{equation}
where $A=St^2{{2e}\over{\hbar}}{{m^*}\over{\pi\hbar^2}}$.
As seen in Fig. 1(a), this expression provides a remarkably good fit to
the experimental data, despite a $\Gamma$ (0.15 mV at 0.31 K) and
applied bias which are comparable to the Fermi energy (0.29 mV).


In Fig. 1(b \& c) we summarize the $T$-dependence of values of $A$ and
$\Gamma$ extracted from the fits of our data to Eq. (1). In general,
the width $\Gamma$ is inversely proportional to the quasiparticle
scattering lifetime. As seen in Fig. 1(c), the fitted
widths for all three carrier concentrations appear to extrapolate to
finite values as $T\rightarrow 0$. Static disorder, mainly from ionized
impurities, is believed to lead to a finite zero-$T$ scattering
lifetime, and hence a finite tunneling peak width \cite{Murphy, Ritchie}.
Our observed decrease in the extrapolated zero-$T$ width with increasing carrier
concentration is likely a result of the improved screening of the
disorder, and is consistent with previous bilayer electron data. The
increase of $\Gamma$ with increasing $T$ in the bilayer electron case
has been attributed to carrier-carrier scattering \cite{Katayama, Murphy, Ritchie}.
Jungwirth and MacDonald \cite{Macdonald2} calculated this
scattering, and found excellent, quantitative agreement with the
electron bilayer data \cite{Murphy}. As shown in Fig. 1(d), this prediction
underestimates the change in the measured width as a function of
temperature for our data. The discrepancy
is possibly due to the very large $r_s$ in our
system \cite{Macdonald2}.

In Fig. 1(b) we observe that the amplitude $A$ is $T$ independent to
within $\pm 2.5\%$ for the three densities investigated here. According
to Eq. (1), the amplitude $A$ is proportional to the square of the tunneling
matrix element $t$, which is intuitively $T$ independent.
In a single-particle picture, $t$ is the hopping
matrix element between the two planes, and is related to the energy
splitting between symmetric and antisymmetric wavefunctions spanning
the two wells, $2t=\Delta_{SAS}$. Using a density-of-states
effective mass $m^*=0.2m_0$ \cite{Mass} , we find $\Delta_{SAS}=$ 8.6 nV for
$p=2.4\times10^{10}$ cm$^{-2}$ and slightly smaller for lower $p$
(see Fig. 1(b)). These $\Delta_{SAS}$ values
are 4 orders of magnitude larger than expected from simple
estimates of $\Delta_{SAS}$ based on the structure of our bilayer
system and the hole band parameters \cite{DSAS2}.

\begin{figure}
\includegraphics{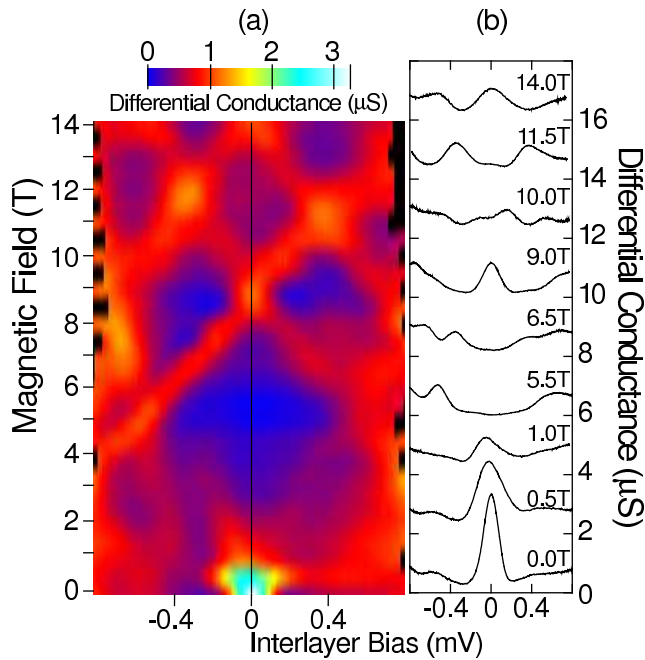}
\caption{\label{Fig4} (a) (Color online) False color plot of the
differential tunnel conductance of the hole bilayer as a function of
interlayer bias and applied parallel magnetic field.
Data were taken at $0.31$K and a hole
density of $2.1\times10^{10}$cm$^{-2}$ per layer. For clarity, the tunneling traces
shown in (b) have been offset vertically (in steps of $2\mu$S).}
\end{figure}

Are strong interlayer correlations responsible for the anomalously
large zero-bias tunneling peak that we observe in our very dilute
bilayer sample? Note that in our sample the ratio of the average
in-plane interparticle distance $\langle r \rangle =1/\sqrt{p}$ to
the interlayer distance $d$ is about 2.9-3.3 (we define $d$ as the
center-to-center distance between the two quantum wells). If the
large tunneling peak is indeed related to the diluteness of our
bilayer layer system, then we would expect a similarly large peak in
samples that have a similar $\langle r \rangle / d$ ratio but whose
barrier is slightly thicker. Our measurements on two other samples
with $\langle r \rangle / d \sim 3$, however, indicate otherwise: In a
sample with a barrier width $w_B = 8.0$ nm we observe a tunneling
peak whose amplitude is a factor of $\sim 10$ smaller than the peak for the
7.5 nm-barrier sample, and in a sample with $w_B = 11$ nm there is
no measurable tunneling. These observations are, on the other hand,
consistent with a single-particle picture in which $\Delta_{SAS}$ is exponentially
sensitive to the width of the barrier \cite{WKB}. To be quantitative, using a self-consistent 
Shroedinger solver for our sample structure, we find that the calculated values of
$\Delta_{SAS}$ approach those extracted from the measured data {\it
for both 7.5 nm- and 8.0 nm-barrier samples} if we take $m_z = 0.19
m_0$ as the mass in the growth direction, 
instead of the expected value $m_z=0.63 m_0$ \cite{DSAS2,VB}.
While our assumption of a single mass for the growth direction 
means we cannot directly interpret this value as being
the mass of holes in AlAs in the growth direction, our data do strongly suggest that 
the hole mass relevant to tunneling through an AlAs barrier
is smaller than has been suggested in the literature \cite{DSAS2}.

The zero-bias tunneling peak arises from a close matching of both
the energy and {\it momentum} of the carriers in the layers. By
applying a magnetic field $B_{||}$ parallel to the layers, the
in-plane (canonical) momenta of the carriers in one well shift with
respect to the other by $k_d=(eB_{||} d)/\hbar$
\cite{LeadbeaterSSC1988}. We would expect the zero-bias peak to have
maximum amplitude at zero field, where the Fermi contours of the two
layers overlap, and decrease rapidly as the carrier momenta in one
layer are shifted with $B_{||}$ and no longer overlap the momenta in
the other layer \cite{EisenFS, Macdonald1}. As shown in Fig.
2, we find that there is indeed a rapid suppression of the
zero-bias peak upon the application of a small $B_{||}$ \cite{BPAR},
of the order of 1 T. However, Fig. 2 reveals a very rich and
puzzling evolution of the tunneling spectra at higher $B_{||}$. If
we assume parabolic in-plane dispersions, the zero-bias tunneling
peak should disappear beyond $B_{||}^c=(2\hbar k_F)/(ed)=2.1$T,
where $k_F=\sqrt{2\pi p}$ is the Fermi wavevector assuming a
spin-degenerate system \cite{Kfermi}. However, well beyond this
field, we find peaks which evolve smoothly as a function of bias and $B_{||}$,
with two dispersing peaks intersecting near zero bias at $\simeq 9$T
and $\simeq 14$T. This suggests a strong distortion
of the hole bands and Fermi contours in a large $B_{||}$. In
particular, a simple change in the band structure curvature, due to
the finite hole layer thickness \cite{GoldoniPRB1993}, would not
explain the re-emergence of a zero-bias peak at high $B_{||}$.

\begin{figure}
\includegraphics{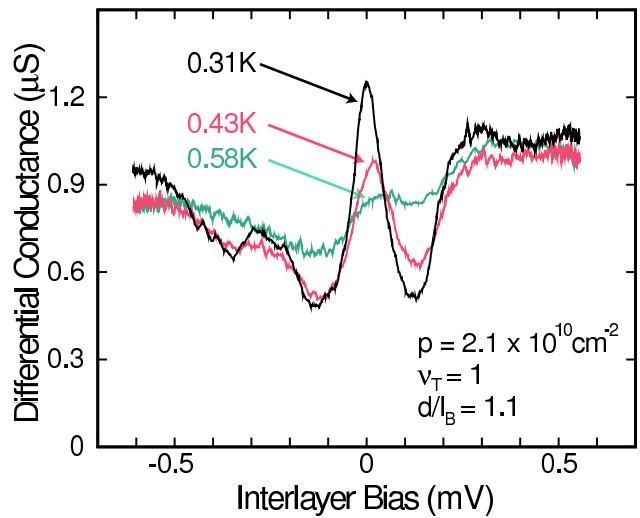}
\caption{\label{Fig5} (Color online) The differential tunnel
conductance as a function of interlayer bias is shown at a layer
density of $2.1\times10^{10}$cm$^{-2}$ in a perpendicular field of $B_{\perp}$=1.72 T,
corresponding to a total (bilayer) filling factor of $\nu_T=1$. }
\end{figure}

Finally we present tunneling data in our bilayer hole system in a
perpendicular magnetic field, $B_{\perp}$. As has been shown in
bilayer electron samples with strong interlayer interaction, when
$B_{\perp}$ is tuned such that the total Landau level filling factor
($\nu_T$) is one, a many-body zero-bias tunneling resonance ensues
\cite{Ian}. With each layer at a filling factor of $1/2$, if the
ratio of the in-plane to the out-of-plane Coulomb energy,
parameterized by $d/l_B$ (where $l_B$ is the magnetic length) is
$\leq 1.8$,  the occupied states in one layer are spatially
correlated with vacant states in the other layer, and tunneling is
resonantly enhanced \cite{Ian}. This pairing has been further
confirmed in both electron and hole bilayers using counterflow
transport studies \cite{Kellogg, Tutuc, Wiersma, Tutuc2}. Here we
provide tunneling data (Fig. 3) for this unique state for our
bilayer 2DHS. Despite a higher background level than the zero-field
tunneling data, there is a strong zero-bias tunneling peak,
persisting up to $T \simeq$ 0.6 K. This is higher than the
temperatures where the equivalent peak in bilayer electron samples
disappears \cite{Spielman, Ian2} and reflects the very strong
pairing in our sample: the parameter $d/l_B=1.1$ for our sample is
indeed smaller than $d/l_B>1.5$ for the published electron bilayer
data. This persistence of the tunneling peak up to higher
temperatures is also consistent with the counterflow measurements in
bilayer hole samples with relatively small $d/l_B$ values
\cite{Tutuc, Tutuc2}. We defer a detailed discussion of the
$\nu_T=1$ bilayer hole tunneling peak height and width on
temperature and $d/l_B$ to a future communication.

In conclusion, we have demonstrated that certain aspects of
interlayer tunneling in dilute bilayer 2D hole systems are
remarkably similar to what has been seen in bilayer electrons. The
shape and width of the zero-field tunneling peak can be well
explained in terms of a Fermi liquid theory developed to understand
the electron data. The anomalously large size of the zero-field peak
is consistent with a smaller than expected hole mass relevant to tunneling through an AlAs barrier. 
The parallel magnetic field dependence of the tunneling
spectra reveals surprises when compared to what is qualitatively
expected based on a simple band structure. Finally, data taken in
perpendicular field show enhanced tunneling at $\nu_T=1$ in bilayer
holes at relatively high temperatures, consistent with the system
being deep in the strongly-interacting regime.

We thank the Princeton NSF MRSEC and DOE for support, and D.A. Huse and R. Winkler for illuminating discussions.

\bibliography{text3}

\begin{thebibliography}{28}
\bibitem[{Bac()}]{Background2}
\bibinfo{note}{In our tunneling measurements we find a small but finite
  background that depends on $p$ and $T$. We also see a small, $T$-dependent
  shift of the tunneling peak from zero bias. We do not know the origin of
  these deviations from the $dI/dV$ expression of Eq. (1). We have included
  them in the fit shown in Fig. 1, and emphasize that they do not affect our
  conclusions.}
\bibitem[{VB()}]{VB}
\bibinfo{note}{The calculated values of $\Delta_{SAS}$ would also approach
  those extracted from the data for both samples if we use $m_z$ = 0.63 $m_0$,
  but take $V_B$ = 0.14 eV instead of the expected value of 0.55 eV. However,
  such a large discrepancy in $V_B$, which is a well-characterized parameter,
  seems unlikely.}
\bibitem[{BPA()}]{BPAR}
\bibinfo{note}{The purely in-plane field was applied at $\sim165^\circ$ from
  the $\langle \bar 233\rangle$ direction and $\sim75^\circ$ from the
  $\langle01\bar1\rangle$ direction.}
\bibitem[{Kfe()}]{Kfermi}
\bibinfo{note}{If we assume only one spin sub-band is occupied, then
  $k_F=\sqrt{4\pi p}$ and $B_{||}^c=3.0$T.}
\end{thebibliography}
\bibliographystyle{apsrev}
\end{document}